\documentclass[anonymous=false, %
               format=acmsmall, %
               screen=true, %
               nonacm=true]{acmart}

\usepackage[ruled]{algorithm2e} 
\usepackage{tikz}
\usetikzlibrary{fit,positioning}

\usepackage{amsmath}
\usepackage{natbib}
\usepackage{xcolor}

\usepackage[utf8]{inputenc}

\DeclareFixedFont{\ttb}{T1}{txtt}{bx}{n}{8} 
\DeclareFixedFont{\ttm}{T1}{txtt}{m}{n}{8}  

\usepackage{color}
\definecolor{deepblue}{rgb}{0,0,0.5}
\definecolor{deepred}{rgb}{0.6,0,0}
\definecolor{deepgreen}{rgb}{0,0.5,0}
\definecolor{deeporange}{rgb}{0.6,0.25,0}
\definecolor{verylightgray}{rgb}{0.97,0.97,0.97}

\usepackage{listings}

\newcommand\pythonstyle{\lstset{
language=Python,
basicstyle=\footnotesize\ttm,
keywordstyle=\ttb\color{deepblue},
commentstyle=\ttm\color{deepgreen},
stringstyle=\color{deeporange},
emphstyle=\ttb\color{deepred},    
emph={MyClass,__init__},          
otherkeywords={self,yield},       
frame=single,                     
showstringspaces=false, 
breaklines=true,
backgroundcolor=\color{verylightgray},
}}

\lstnewenvironment{python}[1][]
{
\pythonstyle
\lstset{#1}
}
{}

\newcommand\pythoninline[1]{{\pythonstyle\lstinline!#1!}}
\renewcommand{\v}[1]{\pythoninline{#1}}

\usepackage{ragged2e}

\title{Joint Distributions for TensorFlow Probability}
\author{Dan Piponi$^\dagger$, Dave Moore$^\dagger$ \& Joshua V.~Dillon}
\thanks{$\dagger${\text{The first two authors contributed equally}}}

\affiliation{%
  \institution{Google Research}
}
\email{tfprobability@tensorflow.org}
\date{December 2019}

\begin{document}

\maketitle

A central tenet of probabilistic programming is that a model is specified exactly once in a canonical representation which is usable by inference algorithms. We describe \v{JointDistribution}s, a family of declarative representations of directed graphical models in TensorFlow Probability. 
\section{Introduction}

TensorFlow Probability \v{JointDistribution}s enable a variety of idioms for probabilistic model specification while providing a standardized interface to inference algorithms. Our design thesis is that there is no single best way to specify joint distributions. Some users may prefer an imperative-style ``probabilistic program,'' while others may prefer a declarative representation of graphical model structure. \v{JointDistribution} invites subclasses to implement different semantics for specifying models, some of which we discuss below. Since subclasses share a common output contract, inference algorithms can access them interchangeably and are isolated from the details of model specification.

While many frameworks separate the low-level distribution abstraction from higher-level probabilistic modeling tools, TFP joint distributions \textit{are} TensorFlow Distributions \citep{dillon2017}: the
\v{JointDistribution} contract is a refinement of the \v{Distribution}
contract. Joint distributions provide \v{sample} and \v{log\_prob} methods, which respectively draw a joint sample and compute a joint log-density. This enables flexible APIs that work seamlessly with both single and joint distributions. Joint distributions naturally extend the concepts of event, batch, and sample shape that organize the Tensor dimensions of \v{Distribution} values. 

Working in TensorFlow offers the potential to exploit hardware acceleration for vectorized sampling and inference. \v{JointDistribution}s support manual and automatic vectorization of models, making it easy to explore new inference paradigms such as massively multi-chain MCMC \citep{hoffman2019langevin} and multi-sample variational bounds \citep{burda2015importance,mnih2016variational,tucker2018doubly}.

\section{The Joint Distribution Interface}

\v{JointDistribution} extends the existing interface of TensorFlow Distributions \citep{dillon2017}. The \v{Distribution} abstraction provides a consistent interface to a large library of probability distributions, of which most are defined over \v{Tensor}s: scalars, vectors, matrices, or batches (replications) thereof. By contrast, a joint distribution is defined over a {\em structure} of \v{Tensor}s: a Python \v{list}, \v{tuple}, \v{dict}, or a nested combination of these with tensors at the leaves. For example, a joint distribution's sample may be a list of tensors, \v{s = [a, b]}, or we may alternatively choose to define the same distribution with dictionary-valued samples, \v{s = \{'a': a, 'b': b\}}. Because hardware-accelerated vectorization is generally most efficient in the SIMD setting where all threads execute the same control flow, our current implementation assumes that the structure itself is the same across all executions.\footnote{The extension to stochastic control flow is interesting future work.}

\subsection{Structured shapes}
\label{sec:shape}

Recall that TensorFlow \v{Distribution} \citep{dillon2017} methods adhere to a ``\v{Tensor}-in, \v{Tensor}-out'' design. Each \v{Distribution} conceptually partitions a \v{Tensor}'s shape into three groups. From left to right: {\em sample shape} indexes independent and identically distributed draws, {\em batch shape} indexes independent but non-identical draws, and {\em event shape} describes a single draw from the underlying distribution. \v{JointDistribution}s extend this design to structures: each \v{Tensor} within the structure is a sample from a conditional distribution, and its shape is interpreted by that distribution under the existing contract.

Just as in \citet{dillon2017}, properties such as \v{dtype}, \v{batch_shape}, \v{event_shape} are unchanging for the lifetime of the joint distribution object. In joint distributions these properties become {\em structures}, matching the structure of the sample.\footnote{Section~\ref{sec:batchshape} discusses some subtleties particular to batch shapes.} For example, where tensor-valued distributions have a \v{dtype}, such as \v{tf.float32}, that specifies the type of their samples, the \v{dtype} of a joint distribution also specifies its {\em structure}; for example, \v{jd.dtype = \{'a': tf.int32, 'b': float64\}} specifies the number of model variables and their names and individual \v{dtype}s, and also that samples are dictionary-valued. 

\subsection{Probabilistic computations}

Many methods of tensor-valued \v{Distribution}s straightforwardly generalize to joint distributions. For a joint distribution \v{jd}, \v{x = jd.sample()} returns a sampled structure of tensors \v{x} in one-to-one correspondence with \v{jd.dtype}. Given a structure \v{x}, \v{jd.log\_prob(x)} returns the \v{Tensor}-valued joint log-density of that structure. Most other probabilistic computations generalize similarly, though not all extend to all joint distributions. For example, joint distributions only implement an analytic mean, standard deviation and entropy in the special case where their components are independent.

Joint distributions also add a new method, \v{sample\_distributions}, which takes a \v{sample\_shape} and optional \v{value}, runs the model forward, returning the resulting structures of sampled conditional distributions \v{ds} and values \v{xs}. This allows users to access finer-grained computations, for example computing every local conditional log-density as \v{[d.log\_prob(x) for (d, x) in zip(ds, xs)]}. The optional \v{value} argument conditions the forward sampling on a partially specified joint value; that is, a structure some of whose leaves are \v{None}. For example, to compute the posterior predictive distributions over observables we might set latent values to posterior samples obtained from inference but leave their children unspecified, to be generated by forward sampling.

Like all TensorFlow distributions, gradients of \v{log\_prob} and \v{sample} (for reparameterizeable distributions) with respect to any model parameters are available by reverse-mode automatic differentiation. Gradient-based inference methods such as Hamiltonian Monte Carlo \citep{neal2011mcmc,hoffman2014no} and black-box variational inference \citep{ranganath2014black,kucukelbir2017automatic} are immediately applicable, although users are of course also free to build custom samplers or more structured variational models.  

\section{Joint Distribution Flavors}

Users construct joint distributions by instantiating one of several subclass `flavors', each of which provides a different interface to a common underlying representation. This `let a thousand flowers bloom' philosophy allows for experimentation in interface design targeting distinct audiences and use cases. In this section we describe three flavors of joint distribution currently implemented in TFP, acknowledging that this is only a small slice of the possibility space.

All flavors described in this paper specify a joint distribution by a sequence of conditional distributions. By the chain rule of probability,
\begin{equation}
\label{chain-rule}
P(X_1, X_2, \ldots, X_n) = P(X_1)P(X_2|X_1)\ldots P(X_n|X_1,\ldots X_{n-1})
\end{equation}
any joint distribution can be written in this form. We leave the exploration of joint distributions specified in other ways for future work.

Under the hood, the sample sites of a joint distribution are addressed using a {\em flattening} of its nested structure into a sequential list representation. Each flavor implements the method \v{ds, xs = jd.\_flat\_sample\_distributions(sample\_shape, values=None)} which returns lists of distributions and values in a canonical order, as well as the methods \v{struct = jd.\_model\_unflatten(list_)} and \v{list_ = jd.\_model\_flatten(struct)} which convert between the canonically ordered internal representation and the user-visible structure. 

\begin{figure}[h]
\noindent\begin{minipage}{0.2\linewidth}
\begin{tikzpicture}
\tikzstyle{main}=[circle, minimum size = 8mm, thick, draw =black!80, node distance = 10mm]
\tikzstyle{connect}=[-latex, thick]
\tikzstyle{box}=[rectangle, draw=black!100]
  \node[main] (m) [] {$m$};
  \node[main, fill = white!100] (s) [above=of m] {$s$};
  \node[main] (x) [right=of m] {$x$};
  \path (m) edge [connect] (x);
  \path (s) edge [connect] (x);
\end{tikzpicture}
\end{minipage}
\begin{minipage}{0.28\linewidth}
\begin{align*}
s &\sim \textsf{InverseGamma}(3, 2) \\
m &\sim \textsf{Normal}(0, 1) \\
x &\sim \textsf{Normal}(m, s)
\end{align*}
\end{minipage}
\begin{minipage}{0.41\linewidth}
\begin{python}[basicstyle=\scriptsize\ttfamily]
simple_model = (
  tfd.JointDistributionSequential([
    tfd.InverseGamma(concentration=3.,
                     scale=2.),    # s
    tfd.Normal(loc=0., scale=1.),  # m
    lambda m, s: tfd.Normal(
      loc=m, scale=s)])            # x
\end{python}
\end{minipage}
\caption{A simple model with conjugate priors on the parameters of a Normal distribution, as a graphical model (left), generative process (center), and \texttt{JointDistributionSequential} (right).}
\label{simple-model}
\end{figure}
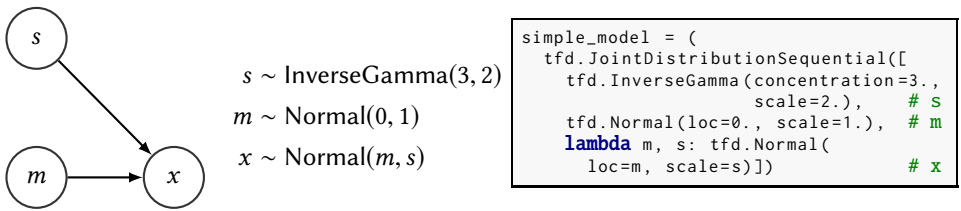

\subsection{JointDistributionSequential}
The first flavor of joint distribution we consider is \v{JointDistributionSequential}.
It is constructed by providing a list of the distributions in the model, using functions to represent distributions conditioned on earlier variables.

Consider the example in Figure~\ref{simple-model}.
Both $s$ and $m$ are independent and are represented by the first two elements in the list.
The variable $x$ is conditioned on both $s$ and $m$, expressed using a \v{lambda} expression with two arguments.
Note that the distributions of $s$ and $m$ are not given names in the code (except in comments).
Arguments to the function are matched with earlier distributions by working up through the list, one element at the time. So the argument \v{m} refers to the previous element and \v{s} refers to the element two before.

On sampling, \v{JointDistributionSequential} draws from each component distribution in turn.
When it encounters a function representing a conditional distribution it provides, as arguments, the appropriate samples generated earlier during the process.

When computing the log-density for a given structure \v{[s, m, x]}, the log-density for each element is computed using the corresponding distribution object.
For conditional distributions, the given elements are provided as arguments to the \v{lambda} expression. This process is carried out for all three component distributions and following Equation~\ref{chain-rule}, the sum of the log-densities is returned.

\subsubsection{Sample shape}
When calling the \v{sample()} method of a \v{JointDistributionSequential} users can also provide a \v{sample\_shape} argument (see Sec. \ref{sec:sampleshape}) to draw multiple samples.
This requires each component distribution in our models to draw multiple samples.
In our example above, the \v{JointDistributionSequential} object supplies the \v{sample\_shape} argument to the distribution for $s$.
Now \v{s} contains a tensor of samples.
Similarly \v{m} will contain a tensor of samples.
Because \v{Normal} treats tensor-valued parameters as a request for a batch of samples (see Section \ref{sec:batchshape}), the distribution for $x$ automatically generates a tensor with a shape matching that of \v{s} and \v{m}.
This means the second normal distribution doesn't need to be provided with a \v{sample\_shape} argument.
The \v{JointDistributionSequential} object can automatically infer that it doesn't need to provide this argument because the third distribution is a distribution \emph{function} rather than a distribution object.

Often we wish to repeatedly draw variables from a distribution \emph{within} a joint distribution, corresponding to plate notation for graphical models. For this we provide the \v{Sample} class that draws multiple samples as a single event.
We also offer the \v{Independent} class to allow the reorganization of batches as individual sample events.

The probabilistic matrix factorization model \citep{mnih2008matrix} demonstrates a nested plate structure.

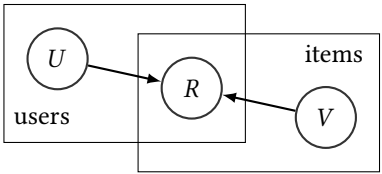
\begin{figure}[h]
\label{matrix-factor}
\noindent\begin{minipage}{0.5\linewidth}
\begin{tikzpicture}
\tikzstyle{main}=[circle, minimum size = 8mm, thick, draw =black!80, node distance = 9mm]
\tikzstyle{connect}=[-latex, thick]
\tikzstyle{box}=[rectangle, draw=black!100]
  \node[main] (R) [] {$R$};
  \node[main, fill = white!100] (U) [above left=-0.2cm and 1.2cm of R] {$U$};
  \node[main] (V) [below right=-0.2cm and 1.2cm of R] {$V$};
  
  \path (U) edge [connect] (R);
  \path (V) edge [connect] (R);
  
  \node[rectangle, inner sep=3mm, draw=black!100,fit= (U) (R),label={[xshift=-11mm,yshift=-17mm]users}] {};
  \node[rectangle, inner sep=3mm, draw=black!100,fit= (V) (R),label={[xshift=10mm,yshift=-5mm]items}] {};
\end{tikzpicture}
\end{minipage}
\begin{minipage}{0.4\linewidth}
\begin{align*}
U_{ij} &\sim \textsf{Normal}(0,\sigma_U) \\
V_{ik} &\sim \textsf{Normal}(0,\sigma_V) \\
R_{jk} &\sim \textsf{Normal}\left(\sum_iU_{ij}V_{ik},\sigma\right)
\end{align*}
\end{minipage}
\vspace{8pt}
\noindent\begin{minipage}{\linewidth}\begin{python}[basicstyle=\scriptsize\ttfamily]
 dist = tfd.JointDistributionSequential(
     [tfd.Sample(tfd.Normal(loc=0., scale=user_trait_scale),  # U
                 sample_shape=[n_factors, n_users]),
      tfd.Sample(tfd.Normal(loc=0., scale=item_trait_scale),  # V
                 sample_shape=[n_factors, n_items]),

      lambda v, u: tfd.Independent(                           # R
            tfd.Normal(loc=tf.matmul(u, v, adjoint_a=True),
                       scale=observation_noise_scale), reinterpreted_batch_ndims=2)])
\end{python}
\end{minipage}
\caption{Probabilistic matrix factorization}
\end{figure}
Both \v{u} and \v{v} are matrices of i.i.d. normal variables so we use \v{Sample} to draw these as single events.
The product of these matrices is a parameter to the third \v{Normal} distribution.
Because this matrix is interpreted as a batch parameter we get a batch of normal variates.
As we wish to treat the entire matrix as a single event we use the \v{Independent} distribution to indicate that the two matrix dimensions should be considered as event dimensions.

\subsection{JointDistributionNamed}
\begin{python}
simple_model = tfd.JointDistributionNamed(dict(
  m=tfd.Normal(loc=0., scale=1.),
  s=tfd.InverseGamma(concentration=3., scale=2.),
  x=lambda m, s: tfd.Normal(loc=m, scale=s)))
\end{python}
When using \v{JointDistributionSequential} the individual distributions are not named and distributions are matched with function arguments implicitly.
\v{JointDistributionNamed} provides explicit naming.
The user provides a dictionary of distributions and functions similar to the list provided to \v{JointDistributionSequential}.
Sampling and log-density computation is also similar, using a topological sort to determine a suitable order to process the distributions.
When sampled, \v{JointDistributionNamed} returns a dictionary of named element samples.

\subsection{JointDistributionCoroutine}
Python provides a type of coroutine \citep{pep342}. A type of function known as a generator, once called, ``yields'' values.
A value yielded by a \v{yield} expression or statement is returned to the caller.
Unlike a standard function, the generator remains in an idle state and the caller can subsequently wake it and transfer control back to it, optionally providing a return value for the \v{yield} expression with which the generator continues.

We use this to provide another way to allow users to write models.
Consider Figure~\ref{simple-model} again:

\begin{python}
def simple():
  Root = tfd.JointDistributionCoroutine.Root
  m = yield Root(tfd.Normal(loc=0., scale=1.))
  s = yield Root(tfd.InverseGamma(concentration=3., scale=2.))
  x = yield tfd.Normal(loc=m, scale=s)
simple_model = tfd.JointDistributionCoroutine(simple_model)
\end{python}
(For now, ignore the \v{Root}s.)
When \v{simple} \v{yield}s its first distribution it isn't immediately assigned to the variable \v{m}.
It is returned to the caller and when the generator is resumed, the caller determines what value is sent back and assigned to \v{m}.

When sampling, the caller samples each yielded distribution and resumes the generator with the sample.
When computing the log-density of a given structure, the log-density for each element is computed according to the yielded distributions.
Instead of sampling the distributions, each given element is provided to the resumed generator.

The \v{JointDistributionSequential} object automatically infers whether a distribution is conditioned on an earlier random variable and uses this to decide when to supply a distribution with a \v{sample_shape} argument.
\v{JointDistributionCoroutine} is unable to make this inference so we use the \v{Root} function to wrap distributions that require a \v{sample_shape} argument.
These correspond to nodes in a graphical model with no parent nodes.

\subsubsection{Latent Dirichlet Allocation}
\begin{figure}[h]
\noindent
\begin{minipage}{0.29\linewidth}
\begin{tikzpicture}
\tikzstyle{main}=[circle, minimum size = 4mm, thick, draw =black!80, node distance = 5mm]
\tikzstyle{connect}=[-latex, thick]
\tikzstyle{box}=[rectangle, draw=black!100]
  \node[main] (theta) [] {$\theta$};
  \node[main] (a) [below=of theta,yshift=-4mm, fill=lightgray] {$\alpha$\vphantom{$\beta$}};
  \node[main] (z) [right=of theta] {$z$};
  \node[main] (w) [right=of z] {$w$};
  \node[main] (b) [below=of w,yshift=-4mm, fill=lightgray] {$\beta$};
  \node[main] (n) [below=of z,yshift=-4.5mm] {$n$\vphantom{$\beta$}};
  \path (n) edge [connect] (z);
  \path (theta) edge [connect] (z);
  \path (z) edge [connect] (w);
  \path (a) edge [connect] (theta);
  \path (b) edge [connect] (w);
  \node[rectangle, inner sep=4mm, draw=black!100,fit= (z) (w),label={[xshift=-7mm,yshift=-4mm]words}] (fred) {};  
  \node[rectangle, inner sep=3mm, draw=black!100,fit= (theta) (z) (w) (fred),label={[xshift=-14mm,yshift=-5mm]topics}] {};  
\end{tikzpicture}
\end{minipage}
\begin{minipage}{0.7\linewidth}
\begin{python}[basicstyle=\scriptsize\ttfamily]
def lda_model():
  Root = tfd.JointDistributionCoroutine.Root
  n = yield Root(tfd.Poisson(rate=avg_doc_length))
  theta = yield Root(tfd.Dirichlet(concentration=
    tfp.util.TransformedVariable(alpha, tfb.Softplus())))
  z = yield tfd.Multinomial(total_count=n, probs=theta)
  w = yield tfd.Independent(
    tfd.Multinomial(total_count=z, logits=tf.Variable(beta)),
    reinterpreted_batch_ndims=1)
lda = tfd.JointDistributionCoroutine(lda_model)
\end{python}
\end{minipage}
\caption{Latent Dirichlet Allocation (LDA) as a plated graph (left) and corresponding \texttt{JointDistributionCoroutine} model (right). Note that $\alpha,\beta$ are learnable via Type-II MLE; see Sec.~\ref{sec:learnParams}.}
\label{lda-model}
\end{figure}
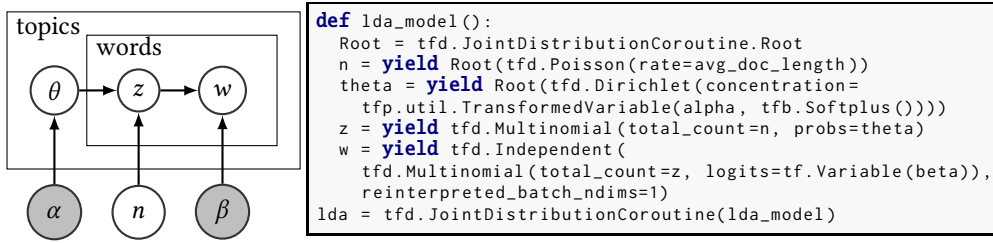

Figure~\ref{lda-model} shows a simplified latent Dirichlet model.
The variable $n$ is the number of words in a document.
If we represent multiple documents as rows of an array, each row would have a different length, requiring a ragged array.
This is not well suited to SIMD parallelism.
Instead, because LDA treats a document as a bag of words, we use $z$ and $w$ to represent the number of occurrences of each topic and word respectively, giving fixed width arrays.

In our implementation using \v{JointDistributionCoroutine} we wrap the distributions for \v{n} and \v{theta} with \v{Root} because they have no parent nodes. When drawing samples for \v{w} we treat \v{z} as a batch of counts so we have one \v{Multinomial} draw for each topic. But we wish the batch to be treated not as separate samples but as the array of counts for a single document, so we use \v{Independent} to group these samples together as a single event.

\section{Vectorized sampling and inference}
\label{sec:sampleshape}
Deep learning frameworks such as TensorFlow \citep{abadi2016tensorflow}, PyTorch \citep{paszke2019pytorch}, and JAX \citep{jax2018github} enable seamless use of vectorized accelerator hardware, such as GPUs, TPUs, and CPU SIMD instruction sets (e.g., SSE, AVX). The ability to run thousands of Markov chains or sample thousands of particles simultaneously opens up new paradigms for inference algorithms and diagnostics (e.g., \citet{hoffman2019langevin}), and to scale probabilistic modeling by vectorizating over batches of data.

Joint distributions support vectorized operations, including sampling and log-density evaluation: \v{jd.sample(sample_shape=N)} returns N independent realizations of the joint distribution, represented as a structure in which each component Tensor has a leading dimension of size N. The log density \v{jd.log\_prob(jd.sample(N))} has shape \v{[N]}, i.e., one joint log-density for each possible world realized in the sample. Vectorized execution of joint distributions enables multiple MCMC chains, multi-sample variational bounds, and multiple optimizations from different initializations. 

\subsection{Batch semantics}
\label{sec:batchshape}

In addition to drawing and evaluating i.i.d. samples as discussed above, it is often valuable to work with a {\em batch} of independent but differently-parameterized distributions. By default, the joint distribution generalization of \v{jd.batch\_shape} is (like other properties in Sec.~\ref{sec:shape}) is just the structure aggregating batch shapes of the individual conditional distributions. Allowing components to have different, but broadcastable, batch shapes can save memory by sharing some local values between multiple (conceptual) batches of the entire joint distribution. This local, per-distribution definition of batch shape also has the interesting consequence that \v{batch\_shape} may depend on \v{sample\_shape}, since i.i.d. samples of root variables will define batches of (non-identical) conditional distributions over their children.

An alternative that can be easier to reason about is to define the batch shape of a joint distribution as a {\em global} property: not a structure, but a single shape describing the number of distinct joint distributions represented. This removes dependence between batch shape and sample shape and can simplify model implementation by eliminating the need to use \v{Independent} to align local batch shapes across components. We have implemented these semantics as part of an experimental set of \v{AutoBatched} joint distributions, which also incorporate automatic vectorization as defined below.

\subsection{Specifying vectorized models}

A common feature of deep PPLs, including Pyro and Edward2 \citep{bingham2019pyro,tran2018simple}, is that exploiting vectorization requires taking care to specify models that vectorize correctly. Consider this innocent-looking \v{JointDistributionCoroutine} model:

\begin{python}
z = yield tfd.Normal(loc=0., scale=[1., 2., 3.])  # z.shape==[3]
x = yield tfd.Normal(loc=0., scale=1.)            # x.shape==[]
y = yield tfd.Normal(loc=z[:2] + x, scale=1.)     # y.shape==[2]
\end{python}

This is a valid probabilistic program; its execution produces a joint sample from $p(z, x, y)$ with shape $([3], [], [2])$. However, this model specification is not valid \textit{under vectorized execution}. We woud expect the shape of $N$ joint samples to be $([N, 3], [N], [N, 2])$, but in fact, given vectorized inputs the \v{z[:2]} on the last line will return a result of shape $[2, 3]$ rather than the $[N, 2]$. Furthermore, neither of those shapes can be added to $x$ having vectorized shape $[N]$ to produce a valid \v{loc}.  Also we should have annotated that 
$z$ and $x$ are root variables that must be sampled $N$ times (as discussed above), while the conditional distribution on the downstream $y$ will already have batch shape $N$ and so needs to be sampled only once.

Joint distributions support two approaches to ensuring that models vectorize correctly. Under \textit{manual vectorization}, the user takes responsibility for specifying a model that is valid under vectorized execution. In this case they would annotate $z$ and $x$ as \v{Root} nodes (discussed above), and apply more careful indexing hints in the last line: \v{tfd.Normal(z[..., :2]) + x[..., None], 1.)}.

Since manual vectorization is often quite cumbersome, joint distributions also support \textit{automatic vectorization} in which the model is treated as a specification of the sampling process for a {\em single} possible world instead of being directly executed with vectorized inputs. This is automated using \v{tf.vectorized\_map}\footnote{Inspired by \v{vmap} in JAX. \citep{jax2018github}} \citep{agarwal2019static} which lifts each tensor operation to one that correctly preserves the batch dimension, allowing na\"ively-written models like the above to execute correctly in parallel. Although manual vectorization may still be preferred when models include operations not supported by \v{vectorized\_map}, or when custom broadcasting allows for additional performance optimizations, we are excited that automatic vectorization can in many cases greatly provide a more user-friendly approach to vectorizing probabilistic models.

\section{Learnable Parameters}\label{sec:learnParams}
All TFP \v{Distribution}s (including \v{JointDistribution}) are trivially "learnable." By "learnable" we mean that the distribution is implicitly a function of some mutable data container (\v{tf.Variable}) which in turn can be updated without necessitating reinstantiation of the \v{Distribution} object. For example,
\begin{python}
loc = tf.Variable(0., name='loc')
scale = tfp.util.TransformedVariable(2., tfp.bijectors.Exp(), name='scale')
jd = tfd.JointDistributionSequential([
  tfd.InverseGamma(concentration=3., scale=scale),
  tfd.Normal(loc=loc, scale=100.)])
jd.trainable_variables
# ==> (<tf.Variable 'scale:0' shape=() dtype=float32, numpy=0.6931472>,
#      <tf.Variable 'loc:0' shape=() dtype=float32, numpy=0.0>)
jd.log_prob([1., 0.])
# ==> <tf.Tensor: shape=(), dtype=float32, numpy=-6.1378145>
loc.assign(-7.)
scale.assign(0.25)
jd.trainable_variables
# ==> (<tf.Variable 'scale:0' shape=() dtype=float32, numpy=-1.3862944>,
#      <tf.Variable 'loc:0' shape=() dtype=float32, numpy=-7.0>)
jd.log_prob([1., 0.])
# ==> <tf.Tensor: shape=(), dtype=float32, numpy=-10.62859>
\end{python}

Notice that accessing \v{jd.trainable_variables} recurses through all dependencies to find \v{tf.Variable}s. The \v{tfp.util.TransformedVariable} serves as a constrained proxy of an otherwise unconstrained \v{tf.Variable} and in this case ensures the scale is non-negative.

All \v{Distribution}s make the contractual commitment to defer reading input arguments until necessity demands. Among other benefits, this enables declaring \v{Distribution}s outside functions which might otherwise update mutable data inputs. For example, maximum likelihood estimation via gradient descent is simply, \pythoninline{tfp.math.minimize(lambda: -tf.reduce_sum(jd.log_prob([p, m])), num_steps, tf.optimizers.Adam(), jd.trainable_variables)}. For do-it-yourself \v{tfp.math.minimize}, see Appendix~\ref{app:diygd}.

\section{Composition of joint distributions}
Since \v{JointDistribution}s are \v{Distribution}s parameterized by \v{Distribution}s, recursive composition of \v{JointDistribution}s (``nesting'') is automatic. For example,
\begin{python}
def _inner():
  Root = tfd.JointDistributionCoroutine.Root
  x0 = yield Root(tfd.Bernoulli(probs=0.25))
  x1 = yield Root(tfd.Normal(loc=0., scale=1.))
jd = tfd.JointDistributionNamed(dict(
  a=tfd.JointDistributionCoroutine(_inner),
  b=tfd.JointDistributionSequential((
    tfd.Poisson(rate=2.),
    tfd.Gamma(concentration=2., rate=1.))),
))
jd.sample()
# ==> {'b': (<tf.Tensor: shape=(), dtype=float32, numpy=1.0>,
#            <tf.Tensor: shape=(), dtype=float32, numpy=0.89002216>),
#      'a': (<tf.Tensor: shape=(), dtype=int32, numpy=1>,
#            <tf.Tensor: shape=(), dtype=float32, numpy=-0.55625236>)}
\end{python}

In Python 3 it is also possible to nest \v{JointDistributionCoroutine} using \v{yield from}.

\section{Related Work}

Like existing deep PPLs such as Pyro \citep{bingham2019pyro} and Edward2 \citep{tran2018simple}, \v{JointDistribution}s enable specifying probabilistic models within a hardware-accelerated differentiable programming environment. Our focus is less on the particular specification language; rather, \v{JointDistribution} is a framework that unifies multiple specification flavors under a shared backend interface, naturally extending the \v{Distribution}s API. Indeed, it would be feasible to build \v{JointDistribution} flavors that use Edward2 or even Pyro (up to TensorFlow-PyTorch translation) as the specification language.

Unlike Edward2, JD flavors have different sample/batch semantics.\footnote{We suspect Pyro also differs from our sample shape propagation design but lacked time to verify this suspicion.} Our insight is to plumb \v{sample_shape} only into the PGM roots and otherwise let shape propagate ``organically'' among descendants. Additionally, some JD flavors provide automatic batching capability.

\v{JointDistribution}s also provide capability similar to effect-handling in Edward2 and Pyro \citep{moore2018effect,phan2019composable}. Sampling from user defined distributions is handled in the \v{JointDistribution} base class where the result is interceptable.

Following its introduction in TFP, PyMC4 adopted a similar interface to \nobreak\hskip0pt\hbox{\v{JointDistributionCoroutine}.} PyMC4 will be built on top of TensorFlow Probability \citep{kochurov2019pymc4}.

Perhaps the most direct analogue to \v{JointDistribution}s in modern PPLs are generative functions in Gen \citep{cusumano2019gen}. Generative functions also support multiple model specification flavors backed by a common interface to inference algorithms. Gen's `choice maps' and the \v{value} structures of joint distributions are roughly analogous. Unlike generative functions, we have not augmented \v{JointDistribution}s with any inherent concept of return values or built-in proposal distributions, and do not currently support models with stochastic control flow. Conversely, joint distributions support efficient vectorized inference, which was not a design goal of Gen. Exploring the connections between these abstractions may be an interesting direction for future development.

\section{Acknowledgements}
We would like to acknowledge the contributions of Junpeng Lao, Alexey Radul, Brian Patton, Matt Hoffman, Christopher Suter,  Pavel Sountsov, and the TFP Team. We also thank Sharad Vikram and Rif A. Saurous for their review of this work.

\bibliographystyle{apalike}
\bibliography{refs}

\appendix
\clearpage
\section{DIY Gradient Descent}\label{app:diygd}

The convenience function \v{tfp.math.minimize} is a wrapper made simple by TF's automatic differentiation. Its (one-step) behavior is roughly equivalent to the following code.

\begin{python}
def make_fit_op(loss_fn, optimizer, trainable_variables):
  @tf.function
  def fit_op(*args, **kwargs):
    with tf.xla.experimental.jit_scope(compile_ops=True):
      with tf.GradientTape(watch_accessed_variables=False) as tape:
        tf.nest.map_structure(tape.watch, trainable_variables)
        loss = loss_fn(*args, **kwargs)
      grads = tape.gradient(loss, trainable_variables)
      optimizer.apply_gradients(list(zip(tf.nest.flatten(grads),
                                         tf.nest.flatten(trainable_variables))))
      return loss
  return fit_op
\end{python}

Which, for example, can be used as:
\begin{python}
def negloglik():
  x, y = next(train_iter)
  return -tf.reduce_sum(model(x).log_prob(y))
fit_op = make_fit_op(negloglik, tf.optimizers.Adam(), model.trainable_variables)
loss = [fit_op() for i in range(num_steps)]
\end{python}

Notice \v{fit_op}'s use of \v{tf.function} and \v{tf.xla.experimental.jit_scope}.  The former ``lifts'' eagerly executed code into graph mode wherein the latter makes numerous optimizations, including op fusion and constant folding.

\end{document}